\documentclass[letterpaper,english,aps,prx,twocolumn,notitlepage]{revtex4-1}
\usepackage{amsmath}
\usepackage{amssymb}
\usepackage{graphicx}
\usepackage[title]{appendix}
\usepackage{color}
\usepackage[normalem]{ulem}
\usepackage{soul}
\usepackage[usenames,dvipsnames,svgnames,table]{xcolor}
\usepackage{dsfont}
\usepackage{cancel}
\usepackage{float}
\usepackage{comment}

\newcommand{\mkdel}[1]{}

\begin{document}
\title{Optimal Control and Glassiness in Quantum Sensing}
\author{Christopher I. Timms}
\affiliation{Department of Physics, University of Texas at Dallas, Richardson, TX, USA}
\author{Michael H. Kolodrubetz}
\affiliation{Department of Physics, University of Texas at Dallas, Richardson, TX, USA}
\begin{abstract}
Quantum systems are powerful detectors with wide-ranging applications from scanning probe microscopy of materials to biomedical imaging. Nitrogen vacancy (NV) centers in diamond, for instance, can be operated as qubits for sensing of magnetic field, temperature, or related signals. By well-designed application of pulse sequences, experiments can filter this signal from environmental noise, allowing extremely sensitive measurements with single NV centers. Recently, optimal control has been used to further improve sensitivity by modification of the pulse sequence, most notably by optimal placement of $\pi$ pulses. Here we consider extending beyond $\pi$ pulses, exploring optimization of a continuous, time-dependent control field. We show that the difficulty of optimizing these protocols can be mapped to the difficulty of finding minimum free energy in a classical frustrated spin system. While most optimizations we consider show autocorrelations of the sensing protocol that grow as a power law -- similar to an Ising spin glass -- the continuous control shows slower logarithmic growth, suggestive of a harder Heisenberg-like glassy landscape.
\end{abstract}
\maketitle

Optimal control has become an increasingly useful avenue for preparing non-trivial quantum states in systems including ultra-cold atoms, trapped ions, quantum mechanical emulators, quantum optics, superconducting qubits, as well as quantum computers~\cite{NielsenChuang,Wineland2005,Timoney2008,Jessen2001,Rosi2013,Koch2004,Reed2012,Spoerl2007,Abdelhafez2020,Bason2012,Frank2016,Wigley2016,Sayrin2011}. More recently, optimal control techniques have been applied to improve quantum sensing~\cite{poggiali2018optimal,Chih2021}, where the problem can be recast from attempting to target a particular state to instead target a family of states that are maximally sensitive to the external field that we wish to detect. Much like quantum control protocols, a wide variety of quantum sensing protocols exist, depending on the sensor (single spin, multiple uncorrelated spins, or entangled spins) and the quantity being measured~\cite{Vandersypen2005,Tosner2009,Liu2017,Zhou2020}. However, from an information theoretic point, the problem remains well-posed: for a given family of control parameters and measurements, maximize the (classical) Fisher information. This leads to minimum sensitivity via the Cramer-Rao theorem~\cite{Yu2022}, providing a measurement protocol that is theoretically optimal.

In this work, we focus on the case of single qubit or ensemble sensors, such as nitrogen vacancy (NV) centers in diamond. Our goal is two-fold. First, we study generalizations of previous protocols in which optimal control was used strictly to determine timing of $\pi$ pulses. Specifically, we consider the generalization to continuously varying magnetic fields and optimization of pump protocols for sensing tasks in which a pump creates the signal that is then measured by the probe. This latter example is relevant to numerous settings, such as NV-sensing of nuclear magnetic resonance signals~\cite{Pham2016,Maze2008} or Nernst photocurrent in transition metal dichalcogenides~\cite{Staudacher2013,DeVience2015,Lovchinsky2017}. Second, we attempt to find a systematic approach for understanding the difficulty of the optimal control problem. By using a mapping between optimal control protocols and (classical) spin systems~\cite{Day2019}, we characterize the optimal control landscape by comparing to that of a spin glass. We find two qualitatively different behaviors: with only $\pi$ pulses, the system behaves like an Ising spin glass, while with continuous control, it behaves similarly to a Heisenberg spin glass. The increased complexity of finding optimal protocols in the Heisenberg landscape is a tradeoff for improved sensitivity, setting a limit on practical applications of optimal control sensing in the lab.

The remainder of the paper proceeds as follows. In Section \ref{sec:single_qubit}, we analyze a single qubit sensor, comparing the $\pi$-pulse optimization from~\cite{poggiali2018optimal} to a generalized optimization where arbitrary time-dependent $x$ and $y$ control fields are allowed, which we refer to as ``continuous drive.'' In Section \ref{sec:glassiness}, we analyze the difficulty of these optimization protocols, placing them on a similar footing via mapping to the landscape of spin glasses. Finally, in Section \ref{sec:conclusions}, we discuss experimental possibilities, generalizations of these results, and conclusions.

\section{Optimal control sensing protocols}
\label{sec:single_qubit}

We begin by discussing a single qubit sensor made from a nitrogen vacancy center. We start our analysis based on \cite{poggiali2018optimal} and then show how these ideas can be extended beyond optimized $\pi$-pulse protocols.

\subsection{Model}

We consider a quantum sensor formed by a negatively charged nitrogen vacancy (NV) center in diamond subject to time-dependent magnetic fields whose amplitude we wish to detect. NV centers are well-studied defects with robust, addressable energy levels. They are primarily used for sensing magnetic field~\cite{Kuwahata2020} and temperature~\cite{Yang2019} due to pristine control, good sensitivity, and small size. We will focus on the case of magnetic field amplitude sensing for a multi-tone target field which can, for example, emulate a biological environment~\cite{Bison2009}. In addition to the target field, there will of course be environmental noise. For NV centers, a major source of magnetic field noise is uncontrolled carbon-13 nuclear spins in the surrounding diamond. 

The NV center electrons form a spin triplet whose magnetic sublevels are split by a static external field. For sensing, we consider the two-level-system (qubit) created by the $S_z=0$ and $S_z=-1$ states, which may be addressed by a direct radio frequency (rf) transition. Such an NV center qubit may be initialized, manipulated, and read out with a high degree of accuracy~\cite{Doherty2013}. 

The sensing protocol that we consider involves three steps: 1) initialize the qubit in the $+x$ direction using a $\pi/2$ pulse, 2) time evolve in the presence of a controllable pulse sequence, and 3) measure $\langle \sigma_x\rangle$. The Hamiltonian for our model is 
\begin{equation}
    H = \frac{1}{2}\left[\gamma B(t) + \xi (t)\right] \sigma_z+\frac{1}{2}\gamma \boldsymbol{B}_c\cdot \boldsymbol{\sigma}
\end{equation}
where $B(t)=b f(t)$ is a field whose amplitude $b$ we wish to measure, $\boldsymbol{B}_c$ is the control field to be optimized, $\gamma=2\pi \times 2.81\times 10^4$ Hz/$\mu$T  is the NV center gyromagnetic ratio, $\xi(t)$ is a noise term with power spectral density $S(\omega)$, and coordinates are chosen such that $\hat{z}$ points along the NV center axis. 

In order for this qubit to perform well as a quantum sensor, its state must be sensitive to the target field, while being minimally affected by the noise. We quantify this by the (classical) Fisher information
\begin{equation}
    F_N = \sum_x \frac{1}{p_N (x|b)}\left(\frac{\partial p_N (x|b)}{\partial b}\right)^2
    \label{eq:Fisher}
\end{equation}
where $p_N (x|b)$ is the probability of measuring state $x\in\{\pm 1\}$ for field $b$ using $N$ repeated measurements. The Fisher information is related to the sensitivity, a quantity that will serve as a cost function that we want to minimize in order to find the optimal protocol, through the equation:
\begin{equation}
    \eta = \frac{\sqrt{\mathbb{T}}}{\sqrt{NF_N}}
    \label{eq:sensitivity}
\end{equation}
where $\mathbb{T} = NT$ with $T$ being the single-shot sensing time.

For direct comparison with \cite{poggiali2018optimal}, we will use a trichromatic target field,
\begin{equation}
    f(t) = \sum_{i=1}^m w_i \mathrm{cos}(2\pi \nu_i t + \alpha_i)
    \label{eq:threetonedrive}
\end{equation}
where $m=3$ is the number of Fourier components, $\nu_i$ is the Fourier mode, and $w_i$ is the amplitude with $\sum_{i=1}^m w_i = 1$. Finally, we take bath spectral function from Poggiali~\cite{PoggialiDissertation}, which is described in further detail in Appendix \ref{PSDAppendix}.

All of the optimization is done on the time-dependent control field $\boldsymbol{B}_c(t)$. Conventionally, sensing protocols involve a series of $\pi$ pulses timed to average away the effect of noise while allowing the qubit rotation due to the target field to add coherently. A simple example that is very effective for monochromatic target fields is the Carr-Purcell pulse sequence, in which $\pi$-pulses are evenly spaced with period $\tau$ to isolate a signal at frequency $(2\tau)^{-1}$. In \cite{poggiali2018optimal}, it was shown that, for more complex target fields, the sensitivity could be drastically improved by modifying the spacing of the $\pi$ pulses using optimal control, which in our model corresponds to the locations of sharp, $\delta$-function peaks of the transverse control field $B_{c,x}$. After discussing these results for context, we show how these protocols can be further improved by allowing arbitrary (discretized) control over the full transverse field $B_{c,x}$ and $B_{c,y}$. 

\subsection{Methods}

\subsubsection{Optimal Control of $\pi$ Pulses}
\label{subsubsec:optimal_control_pi}

As a baseline for improved control protocols, we first consider a variant of the Carr-Purcell method, in which the control field $\boldsymbol{B}_c(t)$ is used to implement a series of $\pi$-pulses around the $x$-axis. As shown in \cite{poggiali2018optimal}, the measured magnetization $\langle \sigma^x \rangle$ is proportional to phase accumulated during the sensing time, giving a sensitivity of
\begin{equation}
    \eta = \frac{e^{\chi(T)}}{\phi(T)}\sqrt{T}
\end{equation}
where $T$ is the total sensing time,
\begin{equation}
    \phi(T) = \int_0^T \gamma f(t) y(t)dt
    \label{eq:magphase}
\end{equation}
is proportional to the phase accumulated from the target field, and 
\begin{equation}
    \chi(T) = \frac{1}{4\pi}\int_{-\infty}^{\infty}d\omega S(\omega)|y(\omega)|^2
    \label{eq:decoherence}
\end{equation}
is the decay of coherence due to noise. Here $y(t)$ is a function that flips sign $1 \leftrightarrow -1$ at the times where $\pi$ pulses occur.

For a given target field $B(t)$, sensing time, and number of $\pi$ pulses, we can optimize the location of the $\pi$ pulses to minimize sensitivity. This is a straightforward optimization problem in a relatively high-dimensional space, which can be approximately solved through a variety of methods~\cite{poggiali2018optimal}. For comparison with other optimization problems, we choose the well-established method of stochastic gradient descent (SGD) throughout this paper~\cite{BaityJesi2018,Day2019}. The stochastic gradient descent algorithm is constructed with the MATLAB language and uses the Adadelta method discussed in Zeiler~\cite{ZeilerAdadelta}. 

The Adadelta method works by having set of parameters, which can be expressed as $\theta_t$ for a certain iteration $t$ of the algorithm, and the gradient of the parameters $g_t$. The goal is to find the appropriate update to $\theta_t$, which is expressed as $\Delta \theta_t$, from $g_t$. In order to do this, it is important to calculate the time dependent learning rate which is dependent upon two quantities. One of these quantities is $\mathrm{RMS}[\Delta \theta]_t$ or the root mean square error of $\Delta\theta_t$. The error or the decaying average of $\Delta\theta^2_t$ is expressed as $\mathrm{E}[\Delta\theta^2]_t$ and is given by

\begin{equation*}
\mathrm{E}[\Delta\theta^2]_t = \gamma \mathrm{E}[\Delta\theta^2]_{t-1}+(1-\gamma)\Delta \theta^2_t
\end{equation*}
where $\gamma$ is a momentum variable that determines the rate at which an added new term contributes to the decaying average. The root mean square of $\mathrm{E}[\Delta\theta^2]_t$ is given by

\begin{equation*}
\mathrm{RMS}[\Delta \theta]_t = \sqrt{\mathrm{E}[\Delta\theta^2]_t+\epsilon}
\end{equation*}
where $\epsilon$ is a very small constant that prevents the quantity from becoming zero. The other quantity used to obtain the learning rate is the root mean square error of $g_t$, where $\mathrm{E}[g^2]_t$ is given by:

\begin{equation*}
\mathrm{E}[g^2]_t = \gamma \mathrm{E}[g^2]_{t-1}+(1-\gamma)g^2_t
\end{equation*}
and the root meas square of $\mathrm{E}[g^2]_t$ is given by the same method for $\mathrm{E}[\Delta\theta^2]_t$. The update for the next iteration of stochastic gradient descent is given by

\begin{equation*}
\Delta \theta_t = -\frac{\mathrm{RMS}[\Delta\theta]_{t-1}}{\mathrm{RMS}[g]_t}g_t
\end{equation*}
and

\begin{equation*}
\theta_{t+1} = \theta_t+\Delta\theta_t
\end{equation*}
When the appropriate parameters are used, the algorithm will be highly effective at descending to the global minimum~\cite{ZeilerAdadelta}.

\subsubsection{Optimal Control with Continuous Drive}

One important question that we address in this work is how generalized control protocols enhance sensing. In particular, while $\pi$-pulses are well-motivated in conventional sensing contexts motivated by their utility in, e.g., measuring nuclear spin precession in NMR, experiments in practice have a much finer degree of control over the transverse magnetic fields used to control the qubit. We consider a relatively simple extension of conventional control protocols in which we allow arbitrary time-dependent control field $\boldsymbol{B}_c(t)$ within the $x$-$y$ plane. This control field is optimized for both $x$ and $y$ rotations separately and the maximum strength for this field allows for a $\pm\pi$ pulse to be implemented for each 50 ns time step that the continuous time evolution is divided into. This extension increases the computational complexity of the control and optimization problem, as the qubit is no longer constrained to precess at the equator. Thus, instead of the phase, we must track the full evolution of the qubit density matrix in the presence of non-trivial bath spectral function $S(\omega)$.

To perform this computation, we begin by discretizing the time evolution into time steps $\Delta t = 50$ ns, which is below the $\pi$-pulse time reported experimentally~\cite{poggiali2018optimal}. During each time step, the Hamiltonian is treated as static with noise term $\xi(t)$ drawn stochastically from the spectral function. For a given noise realization, we compute $\langle x \rangle_\xi$ and $\partial \langle x \rangle_\xi / \partial b$, which can then be averaged over noise realizations to give their values in the noise-averaged density matrix at time $T$. The Fisher information is then 
\begin{equation}
    F_N = \frac{1}{1-\langle x\rangle^2}\left(\frac{\partial \langle x\rangle}{\partial b}\right)^2
\end{equation}
The sensitivity is then given, as before, by Eq.~\ref{eq:sensitivity}. Details on sampling of $\xi(t)$ and calculating $\partial \langle x \rangle_\xi / \partial b$ are much more complicated than for $\pi$ pulses and may be found in Appendix \ref{SensitivityAppendix}.

As with $\pi$ pulses, the optimal protocol for the continuous drive is found through the use of stochastic gradient descent (SGD). However, due to the fact that the implementation of optimal control for continuous drive is very expensive computationally, we found it necessary to improve efficiency by reusing noise configurations during the optimization process. Instead of generating a noise configuration every time the sensitivity or the gradient of the sensitivity is calculated, we begin by generating $20,000$ noise configurations. The gradients associated with the updates of SGD are then calculated using an average over $100$ noise configurations, which are randomly drawn out of the larger dataset of $20,000$. Due to the random nature of this process, this means that at certain stages of the algorithm, one configuration out of the $20,000$ may have been drawn multiple times whereas another configuration may have been drawn zero times.

\begin{figure*}
\centering
	\includegraphics[width=0.75\textwidth]{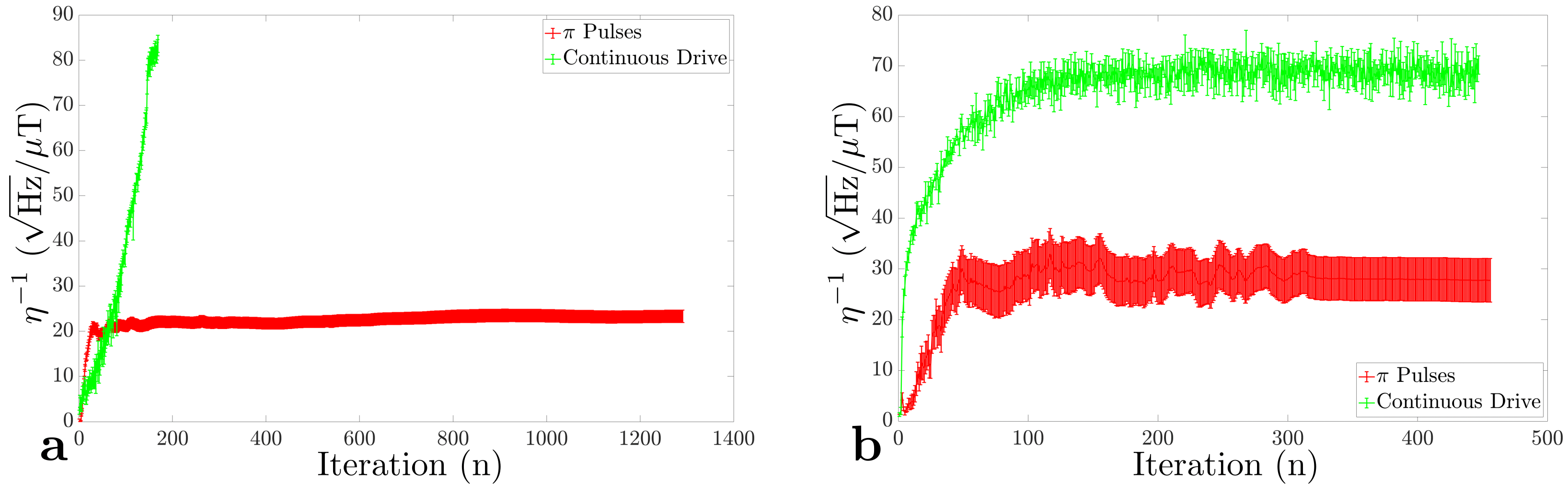}
    \caption{Comparison of the inverse sensitivity for various sensing times using optimal control of $\pi$ pulses and generic time-dependent in-plane fields, a.k.a., continuous drive, as a function of the number of iterations of stochastic gradient descent used to optimize the protocol. While using continuous drive clearly outperforms $\pi$ pulses, this comes at the cost of significantly increased number of iterations before minimum sensitivity is found. Note that we initialize the $\pi$ pulse protocol with evenly spaced $\pi$ pulses as implemented for the CP method, while for the continuous drive each time step is initialized with a random value. These plots use single-shot sensing times of 50 $\mu$s for a) and 100 $\mu$s.}
    \label{fig:Comparison}
\end{figure*}

\subsection{Results}
\label{subsec:single_qubit_results}

Using the methods described in the previous sections, we simulate both $\pi$ pulse and continuous drive evolution for a variety of sensing times and target fields. Figure \ref{fig:Comparison} gives a comparison of the inverse sensitivities achieved for a characteristic choice of target field specified by weights $w = \{0.45, 0.43, 0.12\}$ for three tones at frequencies $\nu = \{77, 96, 144\}$ kHz. The phase factors are $\alpha_i=0.3$ for all three modes. 50 $\pi$ pulses are used to optimize the sensitivity and the target field amplitude is set to $b=1 \mu \mathrm{T}$. It is clear that optimal control with continuous drive performs substantially better than that of $\pi$ pulses, which is not that surprising. One cause  for the improved sensitivity is that using continuous drive allows the qubit to be rotated out of the plane in a controlled fashion. Thus, our optimal control protocol can selectively determine times during the pulse sequence where it is preferable to have the qubit near the equator, for which the phase picked up due to the signal is largest. When the signal is weak, our generalized use of continuous drive allows us to move the qubit away from the equator, so it is much less sensitive to noise.

However, this improved sensitivity comes at the cost of requiring one to solve a much more difficult optimization problem. An initial indication of this can be seen in Figure \ref{fig:Comparison}, in which we clearly see that the number of iterations of gradient descent required to optimize the continuous drive protocol is much greater than the number of iterations for $\pi$ pulses. Partially, this comes from the fact that our search space is much larger. However, as we will see in Section \ref{sec:glassiness}, this increased difficulty is also intricately related to the complexity of the optimal control landscape.

\section{Glassiness of optimal control landscapes} 
\label{sec:glassiness}

It is clear from these simulations that more difficult optimization problems require more gradient descent iterations to converge, but that simple statement is not sufficiently quantitative. In order to more deeply understand the difficulty of these optimization problems, we now discuss connections to other difficult optimization problems, namely equilibration of quenched classical spin glasses.

The connection between glassy physics and optimal control landscapes has been pointed out a number of times in the context of machine learning. Recent results include Baity-Jessy et al.~\cite{BaityJesi2018}, which uses the two-time autocorrelation function, a quantity used to characterize spin-glass physics, to analyze the behavior of neural networks, and Day et al.~\cite{Day2019}, which relates the descent of a protocol, used to optimize fidelity of a system, down the optimal control landscape to that of a spin-glass evolving towards its global minimum after experiencing a quench. 

The connection between glassiness and machine learning/optimal control provides a unifying description allowing us to compare different systems and control protocols. The control protocol plays the role of the spin configuration of the glass. It is described by the $N$-dimensional vector $\boldsymbol{\sigma}$, to be specified more concretely below, where the number of control parameters $N$ is the effective number of spins in the spin glass. Meanwhile, our gradient descent algorithm attempts find $\boldsymbol{\sigma}$ that minimizes sensitivity, similar to how system-bath interactions attempt to push spin glasses into the global minimum of free energy. The number of SGD iterations $n$ therefore plays the role of post-quench time for the spin glass, and we expect glassy phenomena such as aging which reveal properties of the complex landscape.

A useful metric for aging in glassy systems is the two-time autocorrelation function, given by:
\begin{equation}
    \Delta(n_w,n_w+n) = \frac{1}{N} \sum_{i=1}^N (\sigma_i(n_w)-\sigma_i(n_w+n))^2
    \label{eq:twotime}
\end{equation}
which tells us how the system decorrelates between times (SGD iterations) $n_w$ and $n_w+n$. The parameter $n_w$ is referred to as the wait time, and allows us to move away from the short-time transient. In general, $\Delta$ is expected to grow with ``time'' $n$, eventually reaching a plateau in finite systems when a stable minimum is reached. The pre-plateau growth of $\Delta$ can come in many forms, most notably power law growth, $\Delta \sim n^\alpha$, which has been shown to occur in classical Ising spin glasses~\cite{Berthier2004,Komori2000,Parisi1996,Rieger1993}, and logarithmic growth, $\Delta \sim \log n$, which occurs in Heisenberg spin glasses~\cite{Matsumoto2002,Berthier2004}. Therefore, we can look for these dependences as a proxy for the complexity of the optimization landscape; Heisenberg spin glasses are much harder to relax than Ising spin glasses, as indicated by the much slower growth of $\Delta$.

\begin{figure*}
\centering
	\includegraphics[width=0.65\textwidth]{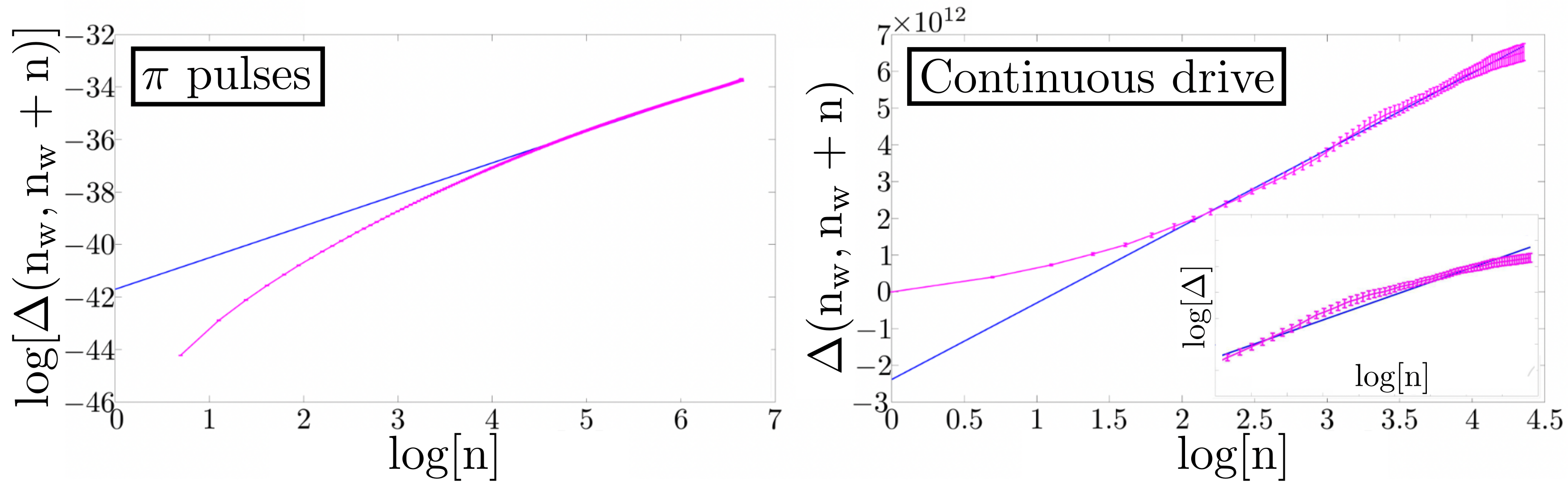}
    \caption{Comparison of two-time autocorrelation function between $\pi$ pulse and continuous drive protocols for single qubit sensing showing consistency with Ising (power law) and Heisenberg (logarithmic) spin glass behavior, respectively. The inset shows the same data on a log-log scale, illustrating that the evolution of the continuous drive protocol is not well-described by power law scaling of $\Delta$. The units of $\Delta$ are $\mathrm{s}^2$ for $\pi$ pulses and $(\frac{\mathrm{rad}}{\mathrm{s}})^2$ for the continuous drive. The parameters used for the calculations using both the $\pi$ pulses and the continuous drive are $w={0.34,0.46,0.20}$ and $\nu={148,55,165}$ kHz and $n_w$ for both cases is 5. The single-shot sensing time is $T=50$ $\mu$s.}
    \label{fig:SpinGlass}
\end{figure*}

We start by examining the single qubit sensing protocols from Section \ref{sec:single_qubit}. An important first step is to choose the mapping from control protocol to effect spin model $\mathbf{\sigma}$. The $\pi$ pulse protocol is determined by the time differences $\Delta t_j = t_j - t_{j-1}$ between the $\pi$ pulses. Therefore, we choose $\mathbf{\sigma} = (\Delta t_1, \Delta t_2, \cdots)$. For continuous drive, the protocol is specified by two vectors of effective magnetic fields (or equivalently phase factors $\Phi$) for the discrete time series, $\Phi_x(t)$ and $\Phi_y(t)$, with $\mathbf{\sigma} = (\Phi_x(0), \Phi_y(0), \Phi_x(\Delta t), \Phi_y(\Delta t), \cdots)$. Autocorrelation functions for these protocols during stochastic gradient descent are shown in Figure \ref{fig:SpinGlass}. For $\pi$ pulses, the data is consistent with a power law scaling $\Delta \sim n^{1.6}$, suggesting that the landscape of optimal control is similar to that of the Ising spin glass. By contrast, for continuous drive, the data is more consistent with a logarithmic scaling, $\Delta \sim 12\log n$, suggesting that the landscape is analogous to a Heisenberg spin glass.

Our results provide a deeper understanding of the relative difficulty of training to find the optimum for $\pi$ pulses versus continuous drive. The results are certainly sensible, as the $\pi$ pulse protocols involves modulation of the signal by a binary ``drive'' term $y(t)$ that alternates between $+1$ and $-1$ depending on the history of $\pi$ pulses applied. Were we to write the protocol in terms of a discretized version of $y$, it would clearly map to an Ising-like problem given the two possible ``spin'' configurations. However, our actual choice of $\boldsymbol{\sigma}$ for defining the protocol is continuous in nature, which is less clearly Ising-like. Since this $\boldsymbol\sigma$ captures the same data as $y(t)$, the scaling of $\Delta$ should be independent of our mapping of the control protocol to an effective spin model. We see similar behavior when optimize an Ising-like (on/off) pump protocol in pump-probe sensing model; see Appendix \ref{sec:photocurrents} for details.

Unlike $\pi$ pulses, continuous drive is described by a planar effective spin, which cannot be reduced to an Ising-like model. Therefore, it is not surprising that the complexity is higher and, as we see, consistent with a Heisenberg spin glass. Due to the difficulty of optimizing for the continuous drive protocol, which require stochastic averaging over many noise realizations, we are unable to evolve to long enough times to clearly see a plateau of $\Delta$. An interesting open question is whether the plateau does indeed form on the same time scale $\log n \approx 4$ as with the $\pi$ pulse case, or whether the increased difficulty manifests as a longer period of growth as expected.

\section{Conclusions}
\label{sec:conclusions}

Using NV center magnetometry as an experimentally relevant testbed, we have shown that generalized pulse sequences and optimal control protocols can dramatically improve sensitivity of quantum sensors. We began by considering optimal control of a single qubit quantum sensor in which pulse sequences are allowed to deviate from conventional $\pi$ pulses used in a wide variety of metrological applications. This continuous drive enables significant improvements in sensitivity, but at the cost of dramatically increased complexity of the optimization problem. We then placed the complexity of these problems on a common ground by considering their connection to classical spin glasses. In particular, we consider the two-time autocorrelation function of the optimal control protocol itself as the system is optimized. We find that the $\pi$ pulse-based protocols undergo power law scaling similar to an Ising spin glass, while the continuous drive protocol is noticeably harder, appearing to satisfy logarithmic scaling characteristic of a Heisenberg spin glass.

Our results have the potential to be immediately used to improve sensitivity in experiments such as measuring Nernst photocurrents in 2D materials~\cite{Zhou2019}. Similar ideas can also be applied to improve other pump-probe sensing protocols, such as optically detected NMR, where very small NMR signals -- created by a series of ``pump'' pulses on the nuclear spins -- are sensed via proximate NV centers whose electrons undergo ``probe'' pulses similar to those we considered. Crucially, we not only test a particular class of experimentally realizable protocols to demonstrate improved sensitivity, but provide an experimentally viable metric for testing how hard the optimization problem is in unknown landscapes, i.e., two-time correlations of the optimal control protocol itself. Experimentalists can, of course, use experimental outcomes to modify their pulse sequence and blindly perform stochastic gradient descent or other similar optimization methods directly. Our analysis suggests a method that can be used to analyze the behavior of the optimization itself at small to moderate number of training steps to understand how far it is useful to push. If, for example, an Ising-like landscape is found, one might be willing to push harder on the optimization protocol than if the problem ends up being Heisenberg-like.

Our work gives rise to a number of important open questions. First, the exponents in the power law remain unexplained. The value of these exponents could help unveil more details of the optimal control landscape. In particular, while there is a sharp distinction between Ising and Heisenberg spin glasses based on their symmetry, within a given class the problem can still vary in difficulty depending on the nature of the effective interactions between the Ising spins and dynamics of the optimization procedure. Second, these ideas can be extended to an arbitrary class of control protocols and optimization techniques, for which the mapping to spin glasses may enable other classes or, potentially, break down completely. A particularly important case is sensing using interacting many-body systems, for which entanglement has been shown to drastically improve the sensitivity~\cite{Zhuang2018,Degen2017}. It will be intriguing to see what is the nature of the optimization landscape in the case of entanglement-enabled sensing and the connection between optimization difficulty and the eventual improvements to sensitivity. Finally, while we have focused on the simplest case of stochastic gradient descent, it remains to be seen how the difficulty of the optimization problem, characterized by the analog with spin-glasses, is affected by utilizing different optimization algorithms, including more advanced forms of machine learning. Here the mapping to spin glass physics may be particularly valuable, as it suggests the use of various methods such as cluster updates which have been used to fundamentally modify the dynamics of solving spin glass problems~\cite{Houdayer2001,Andresen2011}. If a similar technique is used here, it may enable drastic improvements in our ability to optimize the more complicated sensing protocols, with obvious import to experiments in quantum sensing.

\section*{Acknowledgments}

We are grateful for valuable discussions
with Brian Zhou and Murray Holland. This
work was supported by the National Science Foundation through awards
number DMR-1945529 and MPS-2228725 and the Welch Foundation through award number AT-2036-20200401.
Part of this work was performed at the Aspen Center for Physics, which
is supported by National Science Foundation grant PHY-1607611, and
at the Kavli Institute for Theoretical Physics, which is supported
by the National Science Foundation under Grant No. NSF PHY-1748958.
Computational resources used include the Frontera cluster operated
by the Texas Advanced Computing Center at the University of Texas
at Austin and the Ganymede cluster operated by the University of Texas
at Dallas' Cyberinfrastructure \& Research Services Department.

\bibliography{main}

\appendix

\section{Noise spectrum}
\label{PSDAppendix}

\begin{figure}
\centering
	\includegraphics[width=0.9\columnwidth]{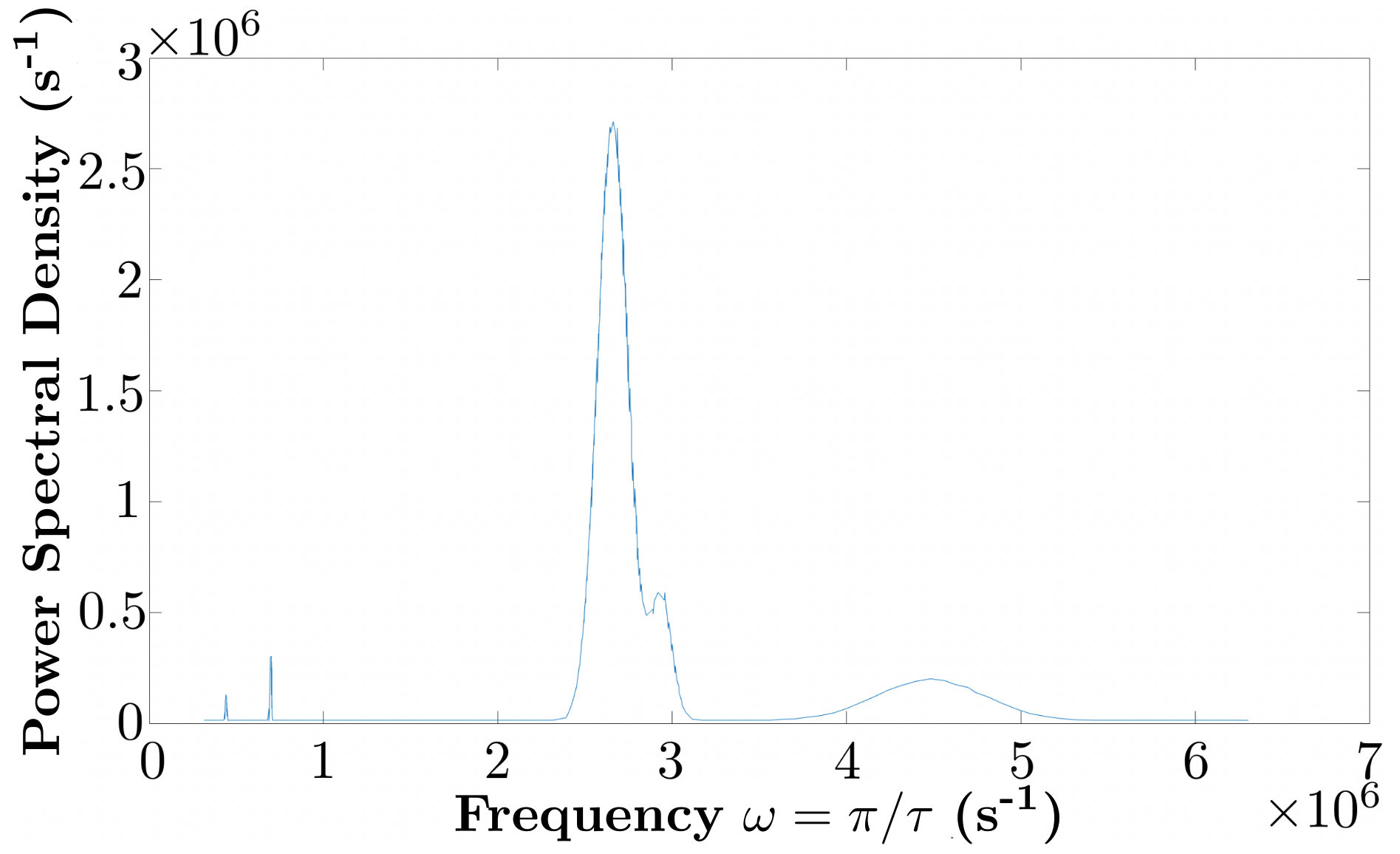}
    \caption{Power spectral density $S(\omega)$ used to calculate the noise that the sensing qubit experiences.}
    \label{PSDPlot}
\end{figure}

The ability to optimize any sensing protocol will depend on the precise form of noise. In this paper, we use the same power spectral density of noise as in \cite{PoggialiDissertation}. This power spectral density $S(\omega)$ is shown in Figure~\ref{PSDPlot}. It is fitted to five Gaussian functions in addition to a constant, whose value is 12170 $\mathrm{Hz}$, that describes the background noise level in this frequency range. The table below displays the Gaussian height, the Gaussian position, and the standard deviation (all in units of $s^{-1}$) for each of these five Gaussian functions.

\begin{center}
\begin{tabular}{||c c c||} 
 \hline
 Gaussian height & Gaussian position & Standard deviation \\ [0.5ex] 
 \hline\hline
 $8\times 10^4$ & $4.407\times 10^5$ & $1 \times 10^4$ \\ 
 \hline
 $2.1377 \times 10^5$ & $6.98 \times 10^5$ & $1 \times 10^4$ \\
 \hline
 $2.6971 \times 10^6$ & $2.6595 \times 10^6$ & $1.2376 \times 10^5$ \\
 \hline
 $5.6288 \times 10^5$ & $2.9353 \times 10^6$ & $8.7174 \times 10^4$ \\
 \hline
 $1.8577 \times 10^5$ & $4.4818 \times 10^6$ & $4.3784 \times 10^5$ \\ [1ex] 
 \hline
 \label{Gaussian}
\end{tabular}
\end{center}

\section{Details of continuous drive}
\label{SensitivityAppendix}

In order to calculate the Fisher information for the case that continuous drive fields are used, we start with a formulation of $p (x|b)$, which is the probability of measuring state $x\in\{\pm 1\}$ for field $b$:
\begin{equation}
    p(x=\pm 1|b) = \frac{1\pm\langle x\rangle_\xi}{2}
\end{equation}
where $\langle x \rangle_\xi$ is the expectation value of measuring the qubit in the $x$-direction averaged over noise realizations $\xi$. Now we incorporate this into the calculation of the Fisher information:
\begin{align}
    F_N&=\sum_x \frac{1}{p(x|b)}\left[\frac{\partial p(x|b)}{\partial b}\right]^2 \nonumber \\ 
    &=\frac{1}{1-\langle x\rangle_\xi^2}\left[\frac{\partial \langle x\rangle_\xi}{\partial b}\right]^2
\end{align}
Now, to calculate $\frac{\partial \langle x\rangle_\xi}{\partial b}$:
\begin{equation}
    \frac{\partial \langle x \rangle_\xi}{\partial b} = \frac{\partial}{\partial b}\mathrm{Tr}\left(\rho_f \hat{x}\right) = \mathrm{Tr}\left(\frac{\partial \rho_f}{\partial b} \hat{x}\right)
\end{equation}
where $\rho_f$ is the density matrix of the qubit after it has been evolved for the entire sensing sequence. The quantity $\frac{\partial \rho_f}{\partial b}$ is calculated using the relation:
\begin{equation}
    \frac{\partial \rho_f}{\partial b} = \mathrm{E}_\xi \left[\frac{\partial U}{\partial b}\rho_i U^\dagger+U\rho_i \frac{\partial U^\dagger}{\partial b}\right]
    \label{delrho}
\end{equation}
where $\rho_i$ is the density matrix of the qubit before it has been time evolved and $U$ is the time evolution matrix for the qubit. The time evolution matrix takes the general form of $U=\tau e^{-i\int_0^T H(t)dt}$ and the Hamiltonian that describes this system is given by:
\begin{equation}
    H(t) = \frac{\gamma b c(t)}{2}\sigma_z+\frac{\xi(t)}{2}\sigma_z+p_x(t)\sigma_x+p_y(t)\sigma_y
    \label{HamApp}
\end{equation}
where $\gamma$ is the gyromagnetic ratio of the NV-center qubit, $b$ is the target field amplitude, $c(t)$ describes the time evolution of the target field, and $p_x(t)$ and $p_y(t)$ describe the control pulses that rotate the qubit in the $x$ and $y$ directions, respectively.

If we decompose the time evolution matrix into a series of time independent unitaries, during which the fields that describe the Hamiltonian in equation~\ref{HamApp} remain constant, as such:
\begin{equation}
    U(T) = U_N U_{N-1}...U_2 U_1
\end{equation}
then it becomes fairly simple to calculate $\frac{\partial U(T)}{\partial b}$:
\begin{multline}
    \frac{\partial U(T)}{\partial b} = U_N U_{N-1}...U_2 \frac{\partial U_1}{\partial b}+U_N U_{N-1}...\frac{\partial U_2}{\partial b} U_1\\+U_N \frac{\partial U_{N-1}}{\partial b}...U_2 U_1+\frac{\partial U_{N}}{\partial b} U_{N-1}...U_2 U_1
\end{multline}
This equation is then incorporated into equation~\ref{delrho}, which can then be used to calculate the Fisher information and the sensitivity. Results are averaged over noise $\xi(t)$, which is sampled in frequency space according to $S(\omega)$ and then  Fourier transformed to give $\xi(t)$.

\section{Ensemble Sensing of Photocurrents}
\label{sec:photocurrents}

As a further test of our methods, in this section we consider NV ensemble sensing of Nernst photocurrents. As recently demonstrated by Zhou et al. ~\cite{Zhou2019}, NV centers are well-designed for studying such material properties, particularly in two-dimensional materials. We will study photocurrent sensing as a canonical example of a pump-probe sensing experiment, as described in further detail below. In the context of optimal control, this has the interesting property that both pump and probe may be optimized, which bears a strong resemblance to other important use cases of NVs for quantum sensing, such a optically detected NMR in which NV centers are used to probe nuclear spins that are, themselves, controlled through standard NMR techniques ~\cite{Boretti2019}.

\subsection{Model}

\begin{figure}
\includegraphics[width=0.7\columnwidth]{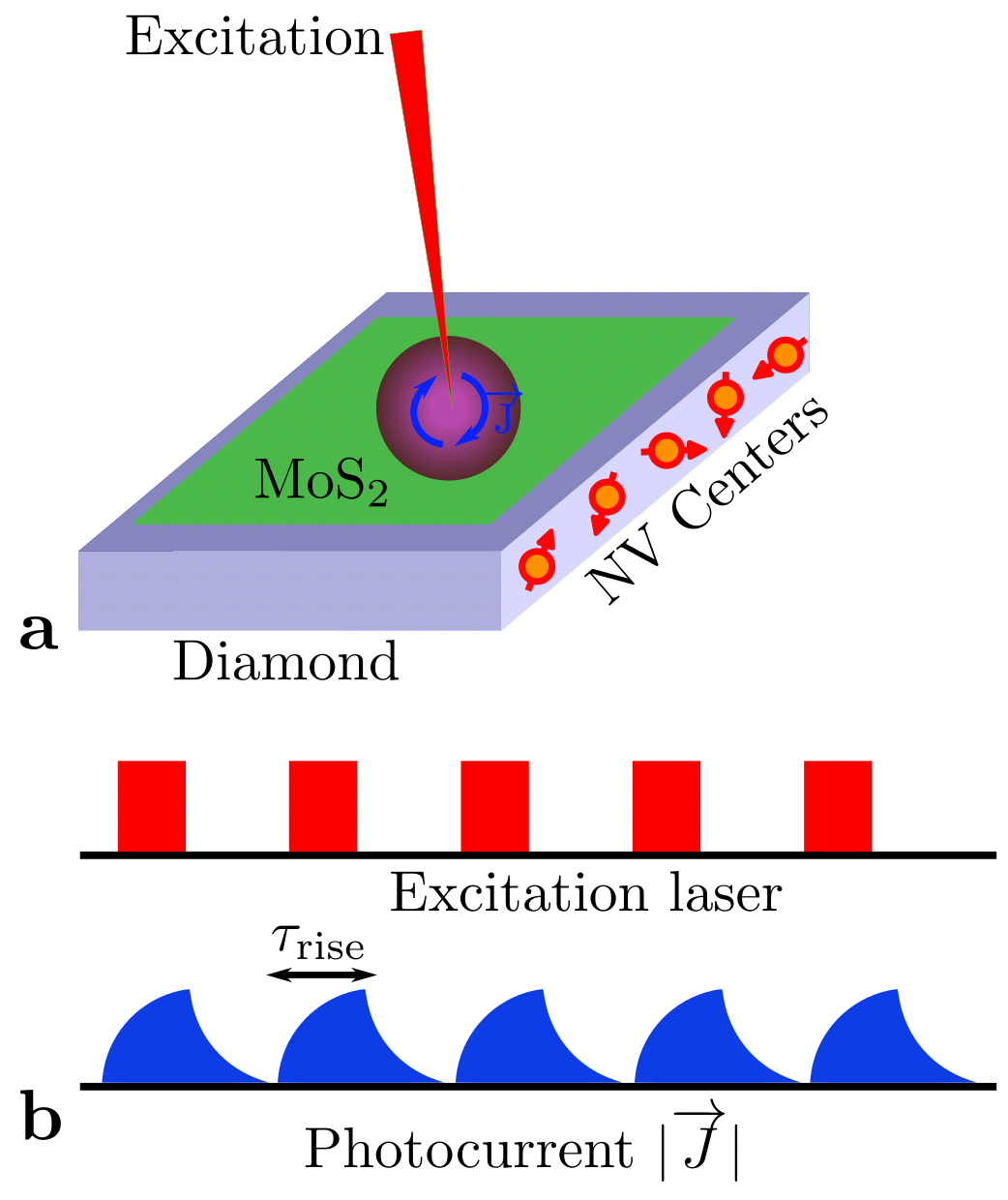}
    \caption{a) Illustration of the experimental setup involved with using NV centers to measure $\mathrm{MoS}_2$ photocurrents. The excitation beam (red) excites the photocurrent. The NV center sensing qubits are located in the diamond represented by the blue sheet, while $\mathrm{MoS}_2$ monolayer is represented by the green two dimensional sheet. The RF pulses rotate the NV center qubits in order to optimize sensing capabilities. b) Time dependence of the excitation beam (top, red), which creates a photocurrent (bottom, blue) whose magnetic field is sensed by the NVs.}
    \label{PhotoExcitation}
\end{figure}

The motivation for this model is ~\cite{Zhou2019}, where evenly spaced $\pi$ (microwave) pulses are used to sense photocurrents in molybdenum disulfide ($\mathrm{MoS}_2$). Figure~\ref{PhotoExcitation} a illustrates the experimental setup. In the experiment, a monolayer of $\mathrm{MoS}_2$ sits on top of a sheet of diamond. Within this diamond, there are nitrogen vacancy centers, which can be used to sense the fields produced by the photocurrent. An excitation beam induces a circular temperature gradient within the $\mathrm{MoS}_2$ monolayer, where the temperature is highest in the center of the circle and becomes lower as the distance from the center of the circle becomes larger. An external magnetic field turns this temperature gradient into a current via the Nernst effect, which is referred to as the Nernst photocurrent. Given the radial dependence of temperature, the current is expected to circulate around the center of the pump laser. Finally, the magnetic field produced by the photocurrent is measured by an ensemble of NV centers, which are operated in a similar regime to what we considered in Section I. A weak probe laser is used to initialize and measure the NV qubits.

Figure~\ref{PhotoExcitation} b illustrates the pulse sequence used in ~\cite{Zhou2019}, in which both the laser pump and microwave control pulses are evenly spaced, in a manner similar to the Carr-Purcell method or a lock-in amplifier. The XY8-N pulse sequence corrects for leading imperfections in the microwave pulses, an effect which we do not treat here but which should be accessible in future iterations. More importantly for us, we model the photocurrent with a finite rise time $\tau_\mathrm{rise}$ in response to turning on/off the pump laser. Similar to Section I, we will consider how this pump-probe sensing protocol can be improved by optimal control over two separate components: the timing of the excitation pulses as well as that of the microwave pulses.

\subsection{Methods}

For a given photocurrent $I_{ph}(t)$ generating magnetic field $B_{ph}(t)$, the NV control and sensing sequence is identical to that in Section I. Therefore, we begin by considering the photocurrent itself. As illustrated in Figure~\ref{PhotoExcitation}, the pump laser is turned ond an off in a series of pulses. Let us denote the times where the laser is switched on/off by $\tau_j$. When the pump laser is turned on at time $\tau_j$, the photocurrent is assumed to exponentially approach a maximum value $B_\mathrm{max} = 0.5016$ mG with time constant
$\tau_\mathrm{rise} = 1.3\ \mu$s:
\begin{equation}
B_{ph}(t) = B_\mathrm{max}-[B_\mathrm{max} - B(\tau_j)]e^{-(t-\tau_j)/\tau_\mathrm{rise}}
\end{equation}
When the pump is turned off, the field exponentially approaches zero on the same time scale. Note that, since the actual value of $B_\mathrm{max}$ drops out of the calculation, we neglect the spatial dependence that gets averaged over in ensemble sensing.

The methods for calculating Fisher information are identical to those considered for $\pi$ pulse optimization of the single qubit, as described in Section I B 1, with the exception that the initial $\pi/2$ (microwave) pulse to place the qubit in the equatorial plane does not take place at $t = 0$ (the time of the first photocurrent laser pulse), but rather a slightly later time $t_0 = 0.897\ \mu$s, which is chosen in line with ~\cite{Zhou2019} to minimize the sensitivity for evenly spaced pulses given the finite rise time $\tau_\mathrm{rise}$. Similar to Section I, the goal is parameter estimation of $B_\mathrm{max}$, the magnitude of the magnetic field generated by the photocurrent. This in turn can be inverted to determine the photocurrent magnitude and connected to the pump laser intensity profile as shown in ~\cite{Zhou2019}. While these initial experiments detected global properties of the photocurrent, next generation devices have the potential to use the spatial locality of the NV centers for measuring the spatial profile of the photocurrent, which could answer some open questions about photocurrent generation in 2D materials.

\subsection{Results}

\begin{figure*}
\includegraphics[width=0.7\textwidth]{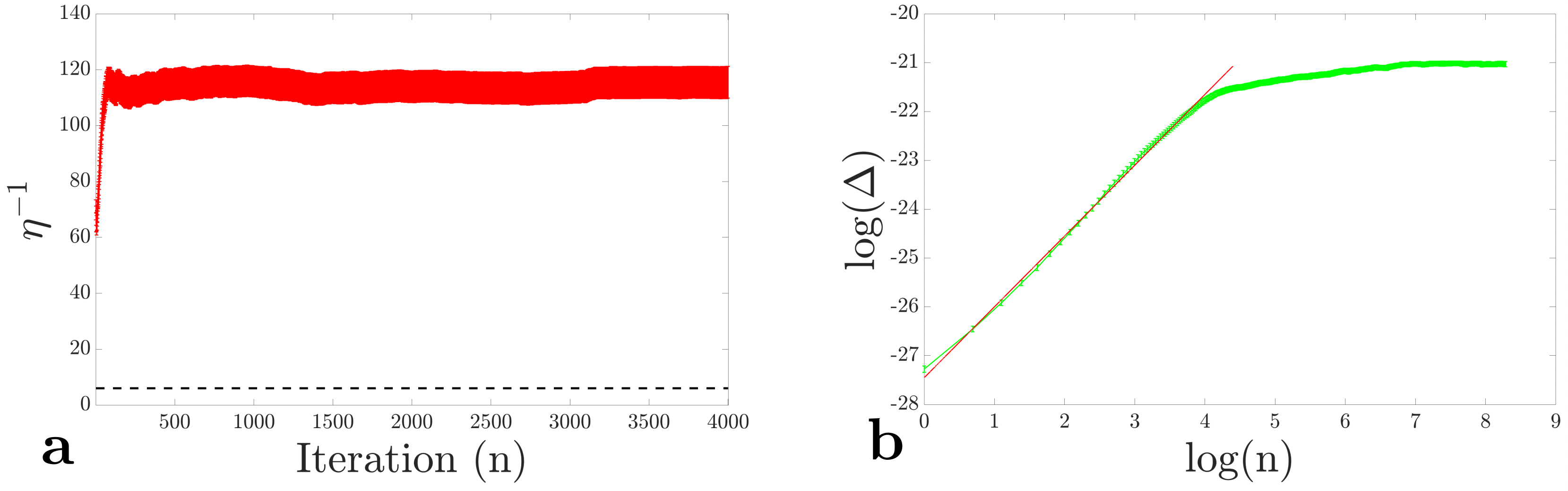}
    \caption{a) Inverse sensitivity during iterations of SGD for optimizing only the pump (photocurrent pulses) with all of the $\pi$ pulses implemented immediately at the start of the sensing sequence (red). Here, the single-shot sensing time ($T$) is $121.6\ \mu$s and the photocurrent pulses are initialized such that they last a time of $\tau = 7.6\ \mu$s and are placed a distance of $\tau$ away from one another. For comparison, the black dotted line shows the case where the microwave pulses are initialized using the CP sequence that starts at time $t_0$ with a $\tau$ of $7.6\ \mu$s with the same initial value of the photocurrent pulses. b) Autocorrelation function for the optimization of photocurrent pulses showing power law fit to $n^{1.45}$. 
    The units $\Delta$ are $s^2$.}
    \label{fig:sensitivity_and_autocorr_photocurrent}
\end{figure*}

Ideally, one would optimize over two parameters: pump (photocurrent pulse) timing and probe (NV $\pi$ pulse) timing. Surprisingly, we found that for our method of modeling the system, this is highly inefficient. In particular, the correlated nature of the optimization makes it such that starting from evenly spaced $\pi$ pulses, as is done the experiments and in the Carr-Purcell sequence is highly non-optimal (see dashed line in Figure \ref{fig:sensitivity_and_autocorr_photocurrent}a). Instead, the optimal control protocol slowly learns to simply eliminate the effect of the $\pi$ pulses by having them all appear at the same time right as the protocol starts (Figure \ref{fig:sensitivity_and_autocorr_photocurrent}a, red). This is likely an artifict of our model, as it fails to capture the lock-in effect used to strengthen signal experimentally by having both pump and probe pulses be evenly spaced.

Nevertheless, even for the case of all $\pi$ pulses up front, the optimization still yields some information about the pump. Figure~\ref{fig:sensitivity_and_autocorr_photocurrent}a shows the optimization of sensitivity for such a case. The magnitude of improvement upon optimizing the photocurrent pulses provides an approximately 27-fold enhancement over the original setup. A large improvement is not terribly surprising due to the fact that photocurrent pulses can be optimized to provide an arbitrary continuously defined phase about the x-axis at any point during the sensing sequence. This gives the photocurrent pulses much more leverage than the optimized sequence of microwave $\pi$ pulses to constructively interfere with the phase acquired by the NV-center qubit due to the target field (the B-field generated by the Nernst photocurrent) while destructively interfering with the noise. 

Finally, for pump optimization, we study glassy autocorrelations. The spin vectors that are used to calculate the two-time autocorrelation function are defined by the timings that determine how the photocurrent pulses are implemented. Given that photocurrent pulses can be mapped to a discretized on/off sequence, we hypothesize that the optimization landscape will be Ising-like. This is consistent with our data in Figure~\ref{fig:sensitivity_and_autocorr_photocurrent}b, where we see apparent power law growth with $\Delta \approx  n^{1.45}$ for the excitation pulses. This power law is less clear than the previous case, as a global minimum is found at rather small n, which is consistent with the relative simplicity of this optimal control landscape. The origin of the power law exponent remains unclear at this time; however, we note that the exponent is quite similar to that obtained for the optimization of $\pi$ pulses during single qubit magnetic field sensing.

\end{document}